\documentclass[fleqn,twoside]{article}
\usepackage{espcrc2}
\usepackage{graphicx}
\usepackage[figuresright]{rotating}

\newcommand{\AmS}{{\protect\the\textfont2
  A\kern-.1667em\lower.5ex\hbox{M}\kern-.125emS}}

\title{The Phase Diagram of High Temperature QCD with Three Flavors of Improved
Staggered Quarks
\thanks{Presented by R.L.~Sugar}}

\author{The MILC Collaboration: C.~Bernard
\address{Department of Physics, Washington University, St.~Louis, MO 63130, USA},
T.~Burch 
\address{Universit\"at Regensburg, Theoretische Physik, 93040 Regensburg, Germany},
C.E.~DeTar
\address{Physics Department, University of Utah, Salt Lake City, UT 84112, USA},
Steven~Gottlieb
\address{Department of Physics, Indiana University, Bloomington, IN 47405, USA},
E.B.~Gregory
\address{Department of Physics, University of Arizona, Tucson, AZ 85721, USA},
U.M.~Heller
\address{American Physical Society, One Research Road, Ridge, NY 11961--9000, USA},
J.E.~Hetrick
\address{Department of Physics, University of the Pacific, Stockton, CA 95211, USA},
R.L.~Sugar
\address{Department of Physics, University of California, Santa Barbara, CA 93106, USA},
D.~Toussaint$\,\null^{\rm e}$
}
       
\begin{document}

\begin{abstract}
We report on progress in our study of high temperature QCD with three flavors
of improved staggered quarks. Simulations are being carried out with
three degenerate quarks with masses less than or equal to the strange
quark mass, $m_s$, and with degenerate up and down quarks with masses in the
range $0.1\, m_s \leq m_{u,d}\leq 0.6\, m_s$, and the strange quark mass fixed
near its physical value. For the quark masses studied to date we find
rapid crossovers, which sharpen as the quark mass is reduced, rather than
bona fide phase transitions.
\vspace{1pc}
\end{abstract}

\maketitle

The MILC Collaboration is studying high temperature QCD with three flavors
of improved staggered quarks~\cite{EARLY} using a one--loop Symanzik improved gauge
action~\cite{GA}  and the Asqtad quark action~\cite{ASQTAD}.  This action
is particularly well suited for the study of high temperature
QCD because it has excellent scaling properties in the lattice spacing,
significantly better dispersion relations for quarks and hadrons than standard
actions, and substantially smaller taste symmetry violations than the
Kogut-Susskind action. 

We are considering two cases:
1) all three quarks have the same mass, $m_q$; and 2) the two lightest
quarks have equal mass, $m_{u,d}$, and the mass of the third
quark is fixed at that of the strange quark, $m_s$. We refer to these cases
as $N_f=3$ and $N_f=2+1$, respectively. This year we have reduced the quark
mass in the $N_f=3$ study to $m_q=0.2\, m_s$, and that in the $N_f=2+1$ study
to $m_{u,d}=0.1\, m_s$. For the masses we have studied to date, we find
rapid crossovers, which sharpen as the quark mass is reduced, rather than 
{\it bona fide} phase transitions. 

We would like to keep the physical quark masses fixed as we vary the
lattice spacing, and therefore the temperature. To this end, for the $N_f=3$
case we work along curves of constant pseudoscalar to vector mass ratio,
$m_{\eta_{ss}}/m_\phi$. These curves are determined from interpolations
between the results of spectrum calculations at lattices spacings 0.12 and 0.20~fm.
For $N_f=2+1$ we make use of spectrum calculations at lattice spacings
0.09, 0.12 and 0.20~fm to determine curves of fixed $m_{\pi}/m_{\rho}$
and $m_{\eta_{ss}}/m_\phi$. Here $\eta_{ss}$ and $\phi$ are the pseudoscalar
and vector mesons made up solely of strange quarks. The lattice spacing,
and hence the temperature, are determined from the heavy quark potential
for values of the gauge coupling and quark masses for which we have
performed spectrum calculations, and by interpolation of the lattice spacing
for other values of the gauge coupling and quark mass.

\begin{figure}
\centerline{\includegraphics[width=2.7in]{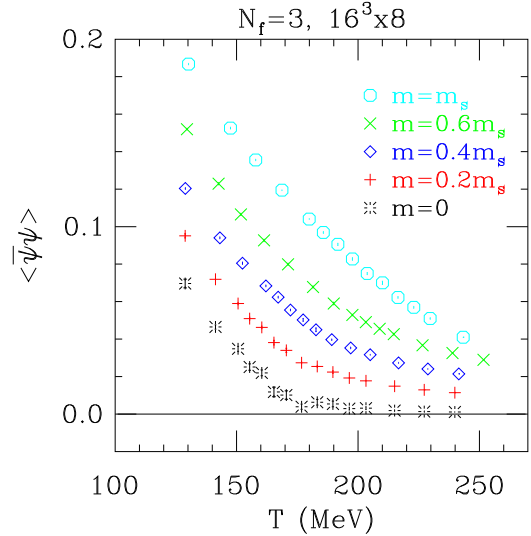}}
\vspace{-7mm}
\caption{The chiral order parameter, $\langle\bar\psi\psi\rangle$, on $16^3\times 8$
lattices for $N_f=3$. The bursts are linear extrapolations of $\langle\bar\psi\psi\rangle$
to $m_q=0$ at fixed temperature.
\label{pbp_nf3_nt8} }
\vspace{-4mm}
\end{figure}

\begin{figure}
\centerline{\includegraphics[width=2.7in]{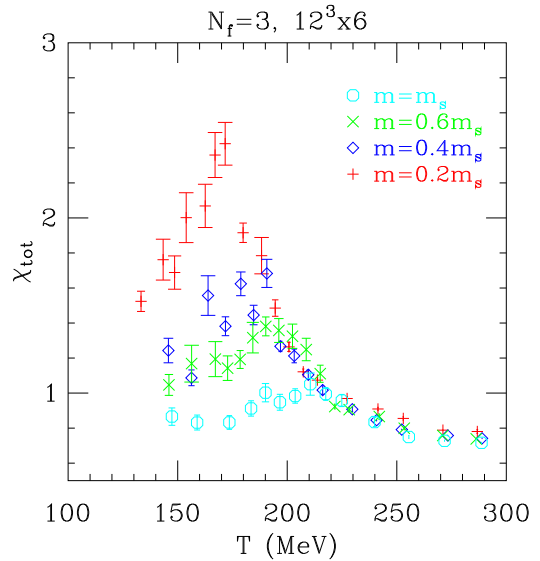}}
\vspace{-7mm}
\caption{The $\bar\psi\psi$ susceptibility as a function of temperature
on $12^3\times 6$ lattices for $N_f=3$. 
\label{chi_tot_nf3_nt6}}
\vspace{-4mm}
\end{figure}

For three equal mass quarks, $N_f=3$, we have carried out thermodynamics studies
on lattices with four, six and eight times slices, and aspect ratio $N_s/N_t=2$.
Here $N_s$ and $N_t$ are the spatial and temporal dimensions of the lattice
in units of the lattice spacing. To date we have studied quark masses in the
range $0.2\, m_s\leq m_q \leq m_s$. In Fig.~\ref{pbp_nf3_nt8} we plot the
chiral order parameter, $\langle\bar\psi\psi\rangle$ as a function of temperature 
on $16^3\times 8$ 
lattices. The bursts are linear extrapolations of $langle\bar\psi\psi\rangle$ from
the runs with the two lightest quark masses 
to $m_q=0$ for fixed temperature. This figure suggests that
for $m_q=0$ there is unlikely to be a phase transition for temperatures 
above 175~MeV, but one could occur at or below that value. For the
quark masses we have studied, the data indicate a crossover, which
sharpens as the quark mass decreases. The absence of a phase transition
at these quark masses is consistent with the findings of the Bielefeld 
group~\cite{BIELEFELD}. The sharpening of the crossover with decreasing
quark mass is seen more clearly in Fig.~\ref{chi_tot_nf3_nt6}, where
we plot the total $\bar\psi\psi$ susceptibility as a function of
temperature on $12^3\times 6$ lattices.

\begin{figure}
\centerline{\includegraphics[width=2.7in]{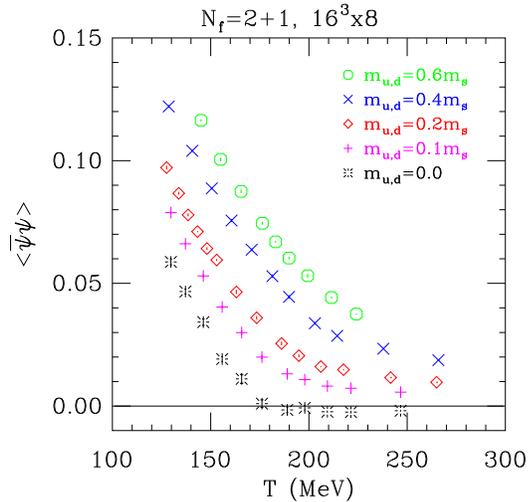}}
\vspace{-7mm}
\caption{The chiral order parameter, $\langle\bar\psi\psi\rangle$, on $16^3\times 8$
lattices for $N_f=2+1$. The bursts are linear extrapolations of $\langle\bar\psi\psi\rangle$
to $m_q=0$ at fixed temperature.
\label{pbp_nf21_nt8} }
\vspace{-4mm}
\end{figure}

The $N_f=2+1$ thermodynamics studies were carried out primarily on $12^3\times 6$
and $16^3\times 8$ lattices. In addition, a number of runs were made on
$18^3\times 6$ lattices to check for finite size effects, which turned out
to be negligible. In this phase of the work we performed simulations with
two degenerate light quarks in the range $0.1\, m_s\leq m_{u,d} \leq 0.6\, m_s$.
In Fig.~\ref{pbp_nf21_nt8} we show the chiral order parameter as a function of 
temperature on $16^3\times 8$ lattices. Here the bursts are a linear extrapolation
from the runs with the two smallest light quark masses to $m_{u,d}=0$ for
fixed temperature and heavy quark mass. This figure indicates that there
is unlikely to be a phase transition for $m_{u,d}=0$ and the heavy quark
mass equal to that of the strange quark for $T> 175$~MeV.  The sharpening 
of the crossover as $m_{u,d}$ decreases is clear.

\begin{figure}
\centerline{\includegraphics[width=2.7in]{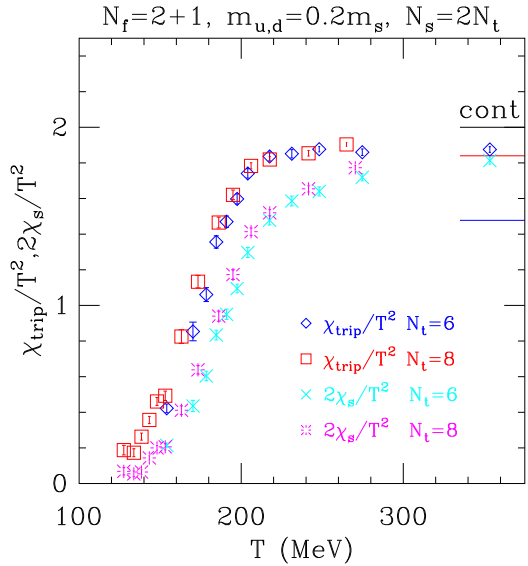}}
\vspace{-7mm}
\caption{The triplet and strange quark number susceptibilities as a function of temperature
for two light quarks with mass $0.2\, m_s$ and one heavy quark with mass $m_s$.
Results are shown
for $12^3\times 6$ and $16^3\times 8$ lattices. The lines on the right of the
figure indicate the values for free quarks in the continuum, and on
the finite lattices on which the simulations were carried out. 
\label{qno_combo} }
\vspace{-4mm}
\end{figure}

Quark number susceptibilities~\cite{SUSC,G_G} are of particular interest 
because they are related to event by event fluctuations in heavy ion collisions,
and can therefore be directly measured~\cite{E_by_E}. They also provide
excellent signals for the crossover.  In Figure~\ref{qno_combo}
we show results for the triplet and strange quark number susceptibilities
on both $12^3\times 6$ and $16^3\times 8$ lattices.
The close agreement between the $N_t=6$ and 8 results illustrates
the excellent scaling properties of the action in the lattice spacing. 

Of particular importance for the coming year will be the determination of
the end point of the line of first order phase transitions expected for
the $N_f=3$ case, and the completion of the $N_f=2+1$ study at $m_{u,d}/m_s =0.1$.

This work is supported by the US National Science Foundation and
Department of Energy and used computer resources at Florida State
University (SP), NCSA, NERSC, NPACI (Michigan), FNAL, and the University of Utah (CHPC).

\end{document}